\documentclass[11pt]{article}

\usepackage[a4paper,margin=2.4cm]{geometry}
\usepackage[T1]{fontenc}
\usepackage[utf8]{inputenc}
\usepackage{lmodern}
\usepackage{microtype}
\usepackage{amsmath,amssymb}
\usepackage{booktabs}
\usepackage{multirow}
\usepackage{array}
\usepackage{algorithm}
\usepackage{algpseudocode}
\usepackage{enumitem}
\setlist{itemsep=2pt, topsep=4pt, parsep=0pt}

\usepackage{xcolor}
\usepackage{tikz}
\usetikzlibrary{positioning,arrows.meta,calc,fit}
\usepackage{pgfplots}
\pgfplotsset{compat=1.17}
\usepackage[numbers,sort&compress]{natbib}
\usepackage{caption}
\usepackage[hidelinks]{hyperref}

\definecolor{dvA}{HTML}{2A78D6}
\definecolor{dvB}{HTML}{EB6834}
\definecolor{dvC}{HTML}{1BAF7A}
\definecolor{dvD}{HTML}{EDA100}
\definecolor{dvE}{HTML}{E87BA4}
\definecolor{dvF}{HTML}{008300}

\title{Streaming Speech Recognition on a Microcontroller:\\
a 65.7M-Parameter Transducer in 2 and 4 Bits on the ESP32-P4}
\author{Billal Boumedine\\[2pt]
\small Faculty of Computer Science, University of Science and Technology Houari Boumediene (USTHB)\\
\small Algiers, Algeria\\
\small \texttt{billal.boumedine@etu.usthb.dz}}
\date{October 2026}

\begin{document}
\maketitle

\begin{abstract}
We run Kroko, a 65.7M-parameter streaming Zipformer2 transducer for English, on the ESP32-P4, a microcontroller with
two 400\,MHz RISC-V cores, a 128-bit SIMD extension (PIE) and 32\,MB of PSRAM, but no neural accelerator. Without any
training, the encoder's 290 matrices are stored with 2 or 4 bits per weight in Hadamard-rotated groups: the width of
each matrix comes from its measured sensitivity, and each matrix is first corrected toward the original model's
output, then rounded with GPTQ and a search over the group norms. The released 71\,MB of int8 weights become a
26.65\,MB file. Activations are int16 between the matrix products and int8 into them. A C engine with hand-written PIE
kernels processes each 1.28\,s chunk of audio in 0.90\,s on both cores, beam search included, and its PC build
computes the same bits as the chip, so accuracy is measured on whole test sets. LibriSpeech WER is 4.34\,\% on
test-clean and 10.26\,\% on test-other, against 3.94\,\% and 9.17\,\% for the released model; on FLEURS and Common
Voice the gap grows to 2.3 and 1.1 points. A GPU simulator shows that the integer arithmetic costs nothing and that
the whole gap comes from the weights, and an ablation on all of test-clean shows that GPTQ does most of the work. We
also report the port's speed history, from 51.3\,s to 0.62\,s of encoder time per chunk, each design step measured
again in one build, an attempt to reach 16\,MB that stopped at 19.63\,MB, and what did not work. We did not measure
energy. The inference engine, the firmware and the 2/4-bit weights are available at
\url{https://github.com/BoumedineBillal/esp32-pie-asr}.
\end{abstract}

\section{Introduction}

Speech recognition on microcontrollers usually means recognizing a fixed list of commands. Espressif's ESP-SR, for
instance, recognizes up to 200 commands on the ESP32-S3~\cite{espsr}. Writing down arbitrary speech needs a model with
tens of millions of parameters, which is more than the memory of most microcontrollers and, until recently, more than
their arithmetic could handle. Oído~\cite{varela2026oido} has just shown a 13M-parameter Conformer-CTC model on the
ESP32-S3 that transcribes in 1.97 times the audio duration on one core: close to real time, not yet in it.

This report describes how we ran a model five times larger, faster than real time and in streaming mode, on the
ESP32-P4. The chip has two 400\,MHz RISC-V cores with a 128-bit SIMD extension called PIE~\cite{esp32p4trm}; it has no
neural accelerator, and the board we used adds 32\,MB of PSRAM and 16\,MB of flash. The model is Kroko~\cite{kroko}, a
streaming Zipformer2 transducer~\cite{yao2024zipformer,graves2012transducer} for English, trained by Banafo and
released as int8 ONNX weights for sherpa-onnx~\cite{sherpaonnx}. We did not retrain it. Everything described here
happens after training: the weights go to 2 and 4 bits, the activations to int16 and int8, and a C engine with
hand-written PIE kernels runs the result.

The system transcribes the board's microphone live and shows the words on the board's screen. On LibriSpeech
test-clean~\cite{panayotov2015librispeech} it reaches 4.34\,\% WER and on test-other 10.26\,\%, with a beam of 4, against
3.94\,\% and 9.17\,\% for the released model run with sherpa-onnx. On real speech from FLEURS and Common Voice the gap
widens to one to two points. Each 1.28\,s of audio takes 0.90\,s to process, and a word becomes final about 2\,s after
it is spoken.

The report makes four contributions.
\begin{itemize}
\item A training-free method that stores a streaming transducer's encoder with 2 or 4 bits per weight in the Cactus
Quants layout~\cite{cactusquants} (Hadamard-rotated groups of 128, Lloyd-Max levels~\cite{lloyd1982,max1960}, one
norm per group): a bit width per matrix from its measured sensitivity, then a sequential pass that corrects each matrix
toward the original model before rounding it with GPTQ~\cite{frantar2023gptq} and choosing the group norms. It brings
71\,MB to 26.65\,MB for half a point of WER on test-clean, and an ablation on the whole test set shows what each part
contributes (Section~\ref{sec:weights}).
\item An integer design, chosen by measurement in a GPU simulator before it was written in C, that costs no measurable
accuracy on LibriSpeech, FLEURS or Common Voice (Sections~\ref{sec:integer} and~\ref{sec:realspeech}).
\item A C engine with PIE SIMD kernels that is bit-exact with its PC build, so that every accuracy number in this
report is measured on complete test sets and is also the chip's, and the cost of each design step measured again in a
single build (Sections~\ref{sec:engine} and~\ref{sec:speed}).
\item An account of what did not work, including an attempt to reach 16\,MB with slimming and quantization-aware
training that stopped at 19.63\,MB (Sections~\ref{sec:push} and~\ref{sec:failed}).
\end{itemize}
The engine and the firmware are open source~\cite{esp32pieasr}, and so are the weights~\cite{hfweights}. The conversion
code is not part of the release; Section~\ref{sec:weights} and Algorithm~\ref{alg:rounding} describe the method in
enough detail to reimplement it.

\section{Related work}

\paragraph{Small speech recognizers.} Whisper tiny.en~\cite{radford2023whisper} (37.8M parameters) and Moonshine
Tiny~\cite{jeffries2024moonshine} (27.1M) are encoder-decoder models that transcribe whole segments; Moonshine was
designed for live transcription on small devices, with compute that scales with the input length. Streaming
transducers~\cite{graves2012transducer} emit tokens as audio arrives. The k2 project's Zipformer
encoders~\cite{yao2024zipformer}, trained with the pruned transducer loss~\cite{kuang2022pruned} in
icefall~\cite{icefall}, are a strong family of them, and sherpa-onnx runs them on CPUs. Kroko belongs to that family.

\paragraph{Speech on microcontrollers.} ESP-SR's MultiNet recognizes a fixed list of commands on the
ESP32-S3~\cite{espsr}. Oído~\cite{varela2026oido} is the closest work to ours. It runs NVIDIA's 13M-parameter
Conformer-CTC model with int8 weights and int8 dynamic activations on one core of the ESP32-S3 (240\,MHz Xtensa LX7, 8\,MB
PSRAM), at 1.97 times the audio duration, and reports 3.70\,\% / 8.23\,\% on test-clean / test-other, measured with a
host build of its engine. A streaming variant with 1.28\,s chunks reaches 4.86\,\% / 11.03\,\% with a 1.3M-parameter
language model (6.27\,\% / 13.06\,\% greedy). Our setting is different: a larger chip, a model five times larger, a
transducer, and streaming as the only mode.

\paragraph{Weight quantization.} GPTQ~\cite{frantar2023gptq} rounds the columns of a matrix one at a time and spreads
each rounding error over the columns not yet rounded, using the inverse of the input's second-moment matrix.
QuIP\#~\cite{tseng2024quipsharp} and QuaRot~\cite{ashkboos2024quarot} rotate weights with Hadamard matrices so that
outliers are spread out and the entries become close to Gaussian; QuIP\# then rounds to lattice codebooks. Cactus
Quants~\cite{cactusquants}, from Cactus Compute, combines a Hadamard rotation per group of 128 with Lloyd-Max levels and
one float16 norm per group. We use that storage layout as documented and contribute the way of rounding into it.
Integer-only inference follows the usual practice~\cite{jacob2018quantization,nagel2021whitepaper}; on
microcontrollers, most published work is on vision~\cite{lin2020mcunet}.

\paragraph{Quantized speech recognizers.} Ding et al.~\cite{ding2022conformer4bit} and Rybakov
et al.~\cite{rybakov2023conformer2bit} report Conformer models with 4-bit and 2-bit weights and no loss of accuracy
on LibriSpeech; both quantize during training. Edge-ASR~\cite{feng2025edgeasr} compares eight post-training methods on
Whisper and Moonshine and finds that 3-bit weights can work on the larger models. Like Edge-ASR we do not retrain, but
most of our weights are at 2 bits, and the result runs on a microcontroller.

\paragraph{Contextual biasing.} For biasing the search toward given words we follow the shallow, in-search approach of Williams et al.~\cite{williams2018contextual}, with a token trie in the manner of
Aho--Corasick~\cite{aho1975}.

\section{Platform and model}
\label{sec:platform}

\paragraph{The chip.} We used Espressif's ESP32-P4X-Function-EV-Board. Its chip is revision v3 of the ESP32-P4 (sold
as ESP32-P4X): two RISC-V cores at 400\,MHz, 768\,KB of internal memory, of which we configure 256\,KB as the L2 cache
for external memory, 32\,MB of PSRAM, here clocked at 250\,MHz, and 16\,MB of flash. The board has an ES8311 codec
with a microphone and a $1024\times600$ MIPI-DSI screen; our chip reports revision v3.2. PIE, the SIMD extension, has eight 128-bit vector registers
and multiply-accumulate instructions that take 16 int8 pairs at once into a wide accumulator register (QACC); 128-bit
loads must be 16-byte aligned, and the low four address bits are silently ignored otherwise. We started the port on
revision v1.3 (PIE 2.1, 360\,MHz) and moved to v3 (PIE 2.2) on 20 October 2026. Our kernels use no instruction that
is new in PIE 2.2, but the encodings changed: 689 of the 1188 instruction words of our kernels differ, and the
kernels ran bit-exact on the first try after a rebuild.

The weights do not fit in the 16\,MB flash, and lossless compression only brought them to about 24.7\,MB. They are
therefore read at boot from a microSD card into PSRAM (2.87\,s at 9.3\,MB/s, after 1.2\,s of card initialization).

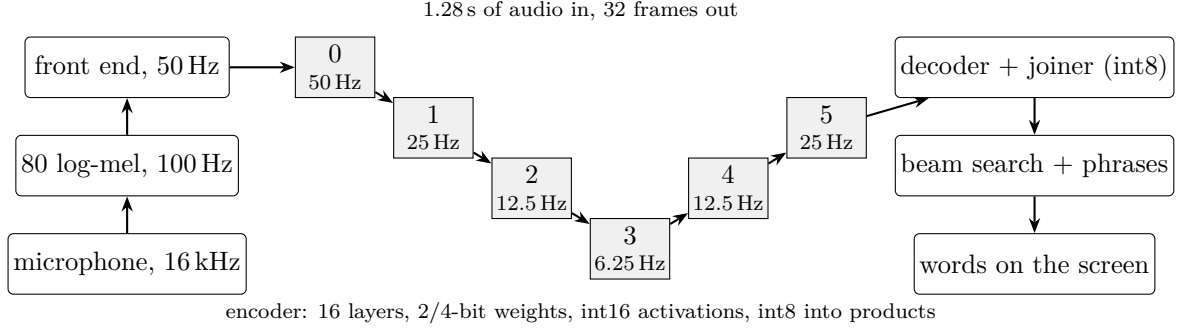
\begin{figure}[tbp]
\centering
\begin{tikzpicture}[
  font=\small,
  box/.style={draw, rounded corners=2pt, minimum width=27mm, minimum height=8mm, align=center, inner sep=2pt},
  st/.style={draw, fill=black!6, minimum width=10.5mm, minimum height=8mm, align=center, inner sep=1pt},
  ar/.style={-{Stealth[length=2mm]}, thick}]
\node[box] (fe) at (0, 0) {front end, 50\,Hz};
\node[box] (fb) at (0, -1.3) {80 log-mel, 100\,Hz};
\node[box] (mic) at (0, -2.6) {microphone, 16\,kHz};
\draw[ar] (mic) -- (fb); \draw[ar] (fb) -- (fe);
\foreach \i/\d/\r in {0/0/50, 1/1/25, 2/2/12.5, 3/3/6.25, 4/2/12.5, 5/1/25}
  \node[st] (s\i) at ({2.75 + 1.3*\i}, {-0.8*\d}) {\i\\[-2pt]\scriptsize \r\,Hz};
\foreach \a/\b in {0/1, 1/2, 2/3, 3/4, 4/5} \draw[ar] (s\a) -- (s\b);
\draw[ar] (fe) -- (s0);
\node[font=\scriptsize, align=center] at (6.0, -3.25)
  {encoder: 16 layers, 2/4-bit weights, int16 activations, int8 into products};
\node[font=\scriptsize] at (6.0, 0.75) {1.28\,s of audio in, 32 frames out};
\node[box] (jn) at (12.0, 0) {decoder + joiner (int8)};
\node[box] (bs) at (12.0, -1.3) {beam search + phrases};
\node[box] (tx) at (12.0, -2.6) {words on the screen};
\draw[ar] (s5) -- (jn); \draw[ar] (jn) -- (bs); \draw[ar] (bs) -- (tx);
\end{tikzpicture}
\caption{The pipeline on the chip. The six encoder stacks run at the frame rates shown, each with 5.12\,s of left
context.}
\label{fig:pipeline}
\end{figure}

\paragraph{The model.} Kroko is icefall's Zipformer2 transducer in its medium configuration: 65.7M parameters, 650 BPE
tokens, a stateless decoder (an embedding of the last two tokens and a grouped convolution) and a joiner. The encoder
has a convolutional front end and six stacks, 16 layers in all, running at frame rates that form a U
(Table~\ref{tab:arch}, Figure~\ref{fig:pipeline}). Each layer has three feed-forward modules, a non-linear attention
module, two self-attention modules sharing one attention-weight computation, two convolution modules and two learned
bypasses. In streaming mode the encoder takes 1.28\,s of audio at a time (128 frames of 80 log-mel features; 141
frames enter with the front end's context) and produces 32 output frames at 25\,Hz. Every stack attends to 5.12\,s of
past frames and keeps its convolution caches between chunks. There is no lookahead, and the state has a fixed size
however long the stream is.

\begin{table}[tbp]
\centering
\caption{Kroko's encoder stacks. Rate: output frames per second. The last column gives the share of each stack's
matrix parameters stored at 4 bits in our 2/4-bit file (Section~\ref{sec:weights}).}
\label{tab:arch}
\small
\setlength{\tabcolsep}{5pt}
\begin{tabular}{rrrrrrrrr}
\toprule
Stack & Layers & Width & Rate (Hz) & Heads & Largest FF & Conv kernel & Matrix params & At 4 bits \\
\midrule
0 & 2 & 192 & 50   & 4 & 512  & 31 & 2.02M  & 69.5\,\% \\
1 & 2 & 256 & 25   & 4 & 768  & 31 & 3.78M  & 40.8\,\% \\
2 & 3 & 384 & 12.5 & 4 & 1024 & 15 & 11.60M & 18.4\,\% \\
3 & 4 & 512 & 6.25 & 8 & 1536 & 15 & 30.22M & 13.8\,\% \\
4 & 3 & 384 & 12.5 & 4 & 1024 & 15 & 11.60M & 100\,\% \\
5 & 2 & 256 & 25   & 4 & 768  & 31 & 3.78M  & 100\,\% \\
\bottomrule
\end{tabular}
\end{table}

\paragraph{Why Kroko.} Before settling on Kroko we compared ready-made streaming Zipformers on four short recordings of
the author's voice (169 words, scored against Whisper large-v3's transcripts). An icefall Zipformer of about the same
size trained on LibriSpeech alone made 50.3\,\% errors on them against 8.9\,\% for Kroko, while it was the better of
the two on a 73-clip LibriSpeech sample (5.39\,\% against 9.41\,\%). The sample is tiny, but the gap is too large to be
noise: with the same architecture and about the same size, the training data decided how well the model handled
speech that is not read audiobooks.

\section{Weights in 2 and 4 bits}
\label{sec:weights}

\subsection{The format}

Take a matrix $W$ with $n$ inputs. Each row is cut into groups of $G=128$ weights ($64$, $32$ or $16$ for the few
widths that are not multiples of 128). A group $w\in\mathbb{R}^G$ is rotated, $u = H_G w$, with $H_G$ the orthonormal
Sylvester--Hadamard matrix, and stored as a float16 norm $\nu$ and $G$ indices $c_k$ of $b\in\{2,4\}$ bits:
\begin{equation}
\hat u_k = \frac{\nu}{\sqrt{G}}\,\ell_{c_k}, \qquad \hat w = H_G\,\hat u,
\end{equation}
where $\ell_0<\dots<\ell_{2^b-1}$ are the Lloyd-Max levels of a standard Gaussian with $2^b$ levels. The cost is
$b+16/G$ bits per weight: 2.125 or 4.125 for groups of 128. The rotation is what makes Gaussian levels fit: after it,
the entries of a group look Gaussian (we measured a kurtosis of 2.97 over Kroko's rotated weights; a Gaussian has 3),
and $\nu$ starts as the group's Euclidean norm, so that $u_k\sqrt{G}/\nu$ has unit variance. A matrix product needs the
input rotated once per group, and that rotation serves every row:
\begin{equation}
y_o = \sum_g \frac{\nu_{o,g}}{\sqrt{G}} \sum_k \ell_{c_{o,g,k}}\,(H_G\,x_g)_k .
\label{eq:product}
\end{equation}
This layout is Cactus Quants~\cite{cactusquants}. What follows, the choice of widths and the rounding, is ours.

\subsection{Sensitivity and the bit plan}

The encoder path has 290 matrices (63.72M parameters): the front end's output projection, the 18 linear maps of each
of the 16 layers, and the joiner's encoder projection. For each matrix alone we rounded it
to 2 bits, then to 4 bits, left everything else as released, and measured the relative error of the encoder output,
\begin{equation}
e = \frac{\sum \lVert \hat y - y\rVert^2}{\sum \lVert y\rVert^2},
\end{equation}
over the frames of held-out calibration clips. The plan starts with every matrix at 2 bits and the small matrices
(groups smaller than 32, 42 of them) at 4. It then moves matrices to 4 bits in decreasing order of error saved per
added byte, $(e_2-e_4)/\text{params}$, while the file stays within a 25\,MB budget. The result keeps 193 of the 290
matrices at 4 bits, 25.35M of the 63.72M parameters.

The plan is not uniform at all (Table~\ref{tab:arch}). The two stacks nearest the output, 4 and 5, end up entirely at
4 bits. Stack 3, the slowest one, which holds 30.22M parameters, close to half the model, has only 13.8\,\% of its
parameters at 4 bits. An earlier study with the same method found the same ordering: the output stacks are the most
sensitive and the middle stack the least, while the single most sensitive matrix is the input projection of the first
layer's self-attention.

\subsection{Rounding}
\label{sec:rounding}

All 290 matrices are rounded one at a time, in the order the forward pass meets them (Algorithm~\ref{alg:rounding}).
Each matrix sees the inputs it receives in the partly rounded model, not in the original one. For a matrix $W$ (rows
are outputs) let $X$ be its inputs in the original model and $X_q$ in the partly rounded model, over all calibration
frames. With $S = X_q^\top X_q$ and $C = X^\top X_q$, three steps follow.

\begin{enumerate}
\item \emph{Correction toward the original model.} Upstream rounding has already changed this matrix's inputs, so we
first solve for the float matrix that best reproduces the \emph{original} output from the \emph{new} inputs,
$\min_{W^*} \lVert X W^\top - X_q W^{*\top}\rVert^2$, with damping:
\begin{equation}
W^* = W + W\,(C - S)\,(S + \lambda I)^{-1}, \qquad \lambda = 0.01\cdot\overline{\operatorname{diag}(S)} .
\end{equation}
When nothing upstream has changed, $C=S$ and $W^*=W$ up to the damping.
\item \emph{Norm search.} In the rotated basis, each group's norm is chosen among ten values, $\nu_0$ times 0.70 to
1.15 in steps of 0.05 (each rounded to float16), to minimize the group's rounding error weighted by
$\operatorname{diag}(S)$; $\nu_0$ is the group's Euclidean norm.
\item \emph{GPTQ.} With the norm fixed, the group's columns are rounded one by one to the nearest level, and each
rounding error is spread over the columns not yet rounded through the Cholesky factor of $S^{-1}$ (with the same
damping), as in GPTQ~\cite{frantar2023gptq}, but in the rotated basis: the rows of $W^*$ are rotated group by group,
and $S$ with them.
\end{enumerate}

\begin{algorithm}[t]
\caption{Sequential rounding into the 2/4-bit format. $\mathcal{H}$ is block-diagonal with one $H_G$ per group;
products with a per-row vector $\nu$ are taken row by row.}
\label{alg:rounding}
\small
\begin{algorithmic}[1]
\Require original model $M$; a bit width $b$ for each encoder-path matrix; calibration frames
\State $\hat M \gets M$
\ForAll{encoder-path matrices $W$ of $\hat M$, in the order of the forward pass}
  \State $X, X_q \gets$ the inputs of $W$ in $M$ and in $\hat M$
  \State $S \gets X_q^\top X_q$; \ $C \gets X^\top X_q$; \ $\lambda \gets 0.01\cdot\overline{\operatorname{diag}(S)}$
  \State $W \gets W + W\,(C-S)\,(S+\lambda I)^{-1}$ \Comment{correction toward $M$}
  \State $V \gets W\mathcal{H}$; \ $S \gets \mathcal{H}^\top S\,\mathcal{H} + \lambda I$; \
         $U \gets$ upper Cholesky factor of $S^{-1}$ \Comment{rotated basis}
  \ForAll{groups $g$ of $G$ columns, left to right}
    \State $\nu \gets$ best of $\nu_0\cdot\{0.70,\dots,1.15\}$ for rounding $V_{:,g}$, error weighted by
           $\operatorname{diag}(S)$ \Comment{norms}
    \ForAll{columns $k$ of $g$}
      \State $c_k \gets$ index of the level nearest to $V_{:,k}\sqrt{G}/\nu$; \ $q \gets \ell_{c_k}\,\nu/\sqrt{G}$
      \State $V_{:,j} \gets V_{:,j} - \bigl((V_{:,k}-q)/U_{kk}\bigr)\,U_{kj}$ for every column $j > k$
             \Comment{GPTQ}
    \EndFor
  \EndFor
  \State store the codes $c$ ($b$ bits each) and the norms $\nu$ (float16); put the rounded $W$ into $\hat M$
\EndFor
\end{algorithmic}
\end{algorithm}

The correction must be done one matrix at a time. Applied to all matrices of a layer at once, it counts the same
upstream error several times, and the encoder error went from 0.034 to 0.33 when we tried it.

\paragraph{Calibration.} The statistics come from 512 clips (87 minutes) of FLEURS~\cite{conneau2023fleurs} English
training audio, chosen in a random order with a fixed seed; no transcripts are used. Clips are batched six at a time,
sorted by length, and padding frames are left out of all statistics. The 24 clips after the first 512 are held out to
measure $e$. Calibrating on the first two hours of the training split instead, which have fewer speakers, made the
result worse on both small test sets we had at the time (13.0 against 10.1\,\% on the author's recordings, 8.13
against 7.27\,\% on the 73-clip sample), so the speaker variety of the calibration audio seems to matter. The
sensitivity pass takes about 12 minutes and the rounding about 25 minutes on a laptop GPU (RTX 4060, 8\,GB).

\paragraph{What is left alone.} The decoder's embedding and projection and the joiner's output layer stay int8, as
released: rounding them to 4 bits alone took the author's recordings from 10.1\,\% to 16.6\,\%. The remaining 1.1M
parameters (biases, norms, depthwise convolutions, gates and bypass scales) are stored in float16.

\begin{table}[tbp]
\centering
\caption{What the weights file holds.}
\label{tab:file}
\small
\begin{tabular}{lr}
\toprule
Part & Parameters \\
\midrule
4 bits: 193 encoder-path matrices & 25.35M \\
2 bits: the other 97 & 38.37M \\
float16: biases, norms, depthwise convolutions, gates & 1.1M \\
int8 (as released): decoder and joiner output & 0.9M \\
\midrule
File on the microSD card & 26.65\,MB \\
\bottomrule
\end{tabular}
\end{table}

\subsection{Ablation}
\label{sec:ablation}

Table~\ref{tab:ablation} takes the method apart on all 2620 clips of LibriSpeech test-clean (53,029 words). Every row
is a file produced by the same code with parts of the method switched off, evaluated in a GPU simulator of the chip's
pipeline with float activations and greedy search; this simulator gives 4.53\,\% for our published 2/4-bit file, and
the chip, with integer activations, 4.52\,\%. The error column is $e$ on the 24 held-out clips.

\begin{table}[tbp]
\centering
\caption{Ablation of the rounding on LibriSpeech test-clean (2620 clips, 53,029 words), float activations, greedy
search, Whisper's English text normalizer. Size: the \texttt{.kq} archive the chip's file is made from (the chip's
file is 0.42\,MB smaller). Error: relative error of the encoder output on 24 held-out FLEURS clips.}
\label{tab:ablation}
\small
\begin{tabular}{lrrrr}
\toprule
Variant & 4-bit matrices & Size (MB) & Error $e$ & WER \\
\midrule
Released weights, unquantized, same simulator & -- & -- & 0 & 4.02\,\% \\
\midrule
Round to nearest, all at 4 bits & 290 & 36.67 & 0.0136 & 4.58\,\% \\
Round to nearest, all at 2 bits (small ones at 4) & 42 & 20.84 & 0.2194 & 21.76\,\% \\
Round to nearest, the 25\,MB plan & 193 & 27.07 & 0.0445 & 8.83\,\% \\
\quad + GPTQ, 128 clips & 193 & 27.07 & 0.0163 & 4.75\,\% \\
\quad + correction toward the original & 193 & 27.07 & 0.0128 & 4.68\,\% \\
\quad + norm search & 193 & 27.07 & 0.0124 & 4.50\,\% \\
\quad + 512 clips: the published file & 193 & 27.07 & 0.0090 & 4.53\,\% \\
\midrule
Full method, all at 2 bits (small ones at 4) & 42 & 20.84 & 0.0557 & 13.40\,\% \\
\bottomrule
\end{tabular}
\end{table}

The reference row comes first. The released weights give 4.02\,\% in the simulator, as in PyTorch (4.02\,\%) and in
sherpa-onnx (4.03\,\%), so the simulator adds no error of its own, and the published file costs 0.51 points: 269
more word errors out of 53,029.

Rounding to the nearest level on the 25\,MB plan nearly doubles the WER, to 8.83\,\%. GPTQ alone does most of the
repair (4.75\,\%). The correction toward the original takes it to 4.68\,\% and the norm search to 4.50\,\%, slightly
better than every matrix at 4 bits rounded to the nearest level (4.58\,\%), a file 36\,\% larger. The last step, 512
calibration clips instead of 128, lowers the encoder error by a quarter (from 0.0124 to 0.0090) and leaves the WER
where it was: 4.53 against 4.50\,\%, a difference of 16 words, well inside the run-to-run spread of about $\pm$30
words that we observed between equivalent configurations. We chose the published file by its error, before running
this ablation; by this measurement, 128 clips would have done as well.

The encoder error is therefore a good guide down to about 0.015 and a poor one below it. From 0.22 to 0.016 it orders
the variants exactly as the WER does, while the four files with an error between 0.009 and 0.014 all land between
4.50\,\% and 4.68\,\%, differences the WER cannot resolve. During development we had seen the same thing: below an
error of about 0.01 the WER stopped moving.

With every large matrix at 2 bits, the full method cuts the WER from 21.76\,\% to 13.40\,\% at 20.84\,MB. That is a
large gain, but not a usable model: the 6.2\,MB that the plan spends on 4-bit matrices is what keeps the WER within
half a point of the original.

\section{Integer arithmetic}
\label{sec:integer}

The first C port computed the products in int8 on PIE and everything between them in float. That glue took 450 to
500\,ms of each 1.28\,s chunk, so the activations had to become integers. We designed them in a batched GPU simulator
of the chip's computation, with 1116 points in the network where a tensor can be left in float or rounded to int8 or
int16, and chose each format by measurement before writing it in C.

\paragraph{Products.} Equation~\eqref{eq:product} maps onto integers as follows. The int16 input group is rotated in
int32 with the unnormalized Hadamard transform, which needs only additions and is exact. The largest magnitude of each
rotated group in each frame then gives a power-of-two shift down to int8. The PIE tile kernel multiplies these int8
values by the int8 levels of 16 rows at a time, accumulating in QACC. The sums are read out with the group's shift
(PIE's read-out takes the shift from a register, so a dynamic scale costs nothing), multiplied by the group's norm in
int16, summed over groups in int32, and returned to int16 with the bias. Dynamic scales matter here: with static int8
scales on the products' inputs, the encoder output was at 26.1\,dB of signal-to-noise ratio against the float
pipeline, and at 37.8\,dB with per-frame scales.

\paragraph{Between products.} Activations are int16 with static scales calibrated on 128 FLEURS clips. The
non-linearities (SwooshL, SwooshR, sigmoid, tanh and the attention's exponential) are tables indexed by the int16
input. BiasNorm sums its squares in integers, residual additions are multiply-and-shift, and the attention's queries,
keys and values, its weights (one power-of-two scale per row) and its caches are int8. The convolution caches stay
int16. The streaming state holds about 815,000 numbers: 0.8\,MB in int8 against 3.3\,MB in float32.

\paragraph{What it costs.} Table~\ref{tab:integer} sums up the measurements. Rounding every activation to int8 is not
an option: on a 73-clip sample, the WER went from 7.27\,\% to 12.75\,\%, while int16 everywhere gave 7.36\,\%. On a
500-clip subset of test-clean (10,634 words), the chosen activations give 3.33\,\%, as in float, and 3.39\,\% once the
products are integer too, six words more. On the whole of test-clean, the float pipeline gives 4.53\,\% and the chip
4.52\,\%; the integer variants we tried along the way landed between 4.50\,\% and 4.59\,\%, a spread of $\pm$20 to 30
words, which is the noise level of the comparison. The same holds on harder speech (Section~\ref{sec:realspeech}):
17.78 against 17.75\,\% on FLEURS and 20.94 against 20.72\,\% on Common Voice.
Measured layer by layer on 20 clips, the signal-to-noise ratio of the integer pipeline against the float one is
27.8\,dB after the first layer, 33.9\,dB at the output of stack 3, and 33.2\,dB at the encoder's output: the error
does not grow with depth.

\begin{table}[tbp]
\centering
\caption{The cost of integer activations. Every row uses the 2/4-bit weights; greedy search, Whisper's normalizer.
Float: float activations in the GPU simulator. The chip: the integer pipeline, through the PC build. The 73-clip
sample is from LibriSpeech dev-clean.}
\label{tab:integer}
\small
\begin{tabular}{llr}
\toprule
Clips & Activations & WER \\
\midrule
\multirow{3}{*}{73 (dev-clean)} & float & 7.27\,\% \\
 & int16 everywhere, static scales & 7.36\,\% \\
 & int8 everywhere, static scales & 12.75\,\% \\
\midrule
\multirow{3}{*}{test-clean, 500} & float & 3.33\,\% \\
 & the chosen activations, float products & 3.33\,\% \\
 & the chosen activations, integer products & 3.39\,\% \\
\midrule
\multirow{2}{*}{test-clean, all 2620} & float & 4.53\,\% \\
 & the chip & 4.52\,\% \\
\midrule
\multirow{2}{*}{FLEURS test, 647} & float & 17.78\,\% \\
 & the chip & 17.75\,\% \\
\midrule
\multirow{2}{*}{Common Voice test, 1000} & float & 20.94\,\% \\
 & the chip & 20.72\,\% \\
\bottomrule
\end{tabular}
\end{table}

\section{The engine}
\label{sec:engine}

\subsection{Kernels}

The engine is C with PIE kernels in assembly (Table~\ref{tab:kernels}); each kernel has a C reference next to it.
Decoding the weights is the part that does not map onto PIE directly: Lloyd-Max levels are not evenly spaced, so the
codes cannot be expanded with shifts. The decode kernels use gather loads from a table that holds, for each code byte,
the int8 levels of its four 2-bit (or two 4-bit) codes. Decoding 2048 two-bit weights takes 851 cycles. The tile
kernel then multiplies 16 decoded rows by one input group for each frame, 168 cycles per frame and $16\times128$ tile,
about 12 multiply-accumulates per cycle.

\begin{table}[tbp]
\centering
\caption{The PIE kernels.}
\label{tab:kernels}
\small
\begin{tabular}{>{\raggedright\arraybackslash}p{0.37\textwidth}>{\raggedright\arraybackslash}p{0.57\textwidth}}
\toprule
Kernel & What it does \\
\midrule
\texttt{pq\_decode2}, \texttt{pq\_decode4} & weight codes to int8 levels through a gather table \\
\texttt{pq\_tile16}, \texttt{pq\_tile16\_n} & 16 rows $\times$ one group $\times$ $T$ frames; the second also applies the norms and shifts \\
\texttt{pq\_fwht} & the integer Hadamard transform of an input group (radix 8) \\
\texttt{pq\_q8\_shift} & int32 to int8 by a power-of-two shift \\
\texttt{pq\_out\_tab16}, \texttt{pq\_out\_add16}, \texttt{pq\_out\_act8} & read-out: activation table, residual add, int8 output \\
\texttt{pq\_sumsq16}, \texttt{pq\_scale\_bypass16} & BiasNorm's sum of squares; its scale fused with the bypass \\
\texttt{pq\_dw16}, \texttt{pq\_dwcomb16} & the convolution modules' depthwise convolution and its combine \\
\texttt{pq\_mul\_q8}, \texttt{pq\_mul16}, \texttt{pq\_lane\_q8}, \texttt{pq\_lane\_or16} & gating products and per-channel scaling \\
\bottomrule
\end{tabular}
\end{table}

\paragraph{Two cores.} The products are shared between the cores: the input frames are quantized half on each core,
then the tiles are split by blocks of output rows, and each core writes only its own outputs. Two cores make the
encoder 1.58 times faster than one (Section~\ref{sec:speed}), not 2: in an earlier measurement, decoding on each core
took 60\,\% longer and the multiply-accumulates 33\,\% longer when both cores ran, because they share the memory.

\paragraph{Memory.} The 26.65\,MB of weights are used in place in PSRAM and read once per chunk, in order. The L2
cache is set to 256\,KB with 128-byte lines, and the L1 and L2 caches fetch the next line on a miss (autoload);
without the autoload the encoder is 45.6\,ms slower. In an earlier build, decoding fake codes instead of reading them
from PSRAM saved 39\,ms of the 79\,ms of decoding per chunk: that is what the PSRAM stream costs, and the most a DMA
prefetch could save. About 20\,KB of hot code (the products' drivers, the attention and first-convolution loops, the
kernels) runs from internal RAM, which saved 40\,ms per chunk when it was introduced.

\subsection{Bit-exact with the PC}

The engine also builds for a PC, with each kernel replaced by its C reference. Three rules make the PC build compute
exactly what the chip computes.
First, every transcendental function comes from a small header built from IEEE addition, subtraction, multiplication
and division only. The C libraries of the two sides (glibc and newlib) differ in the last bit of \texttt{exp},
\texttt{log}, \texttt{tanh} and others, and those bits change the features and the activation tables. (The chip has no
double-precision unit, so building the tables this way made initialization 0.4\,s longer.)
Second, both sides compile with \texttt{-ffp-contract=off}, so no compiler fuses a multiply and an add on one side only.
Third, every kernel has a C reference with the same rounding, and each two-core part writes only its own outputs, so
the order of execution cannot change a result.

We check this on the board: all 22 kernel tests against their references; a CRC of the encoder output after every chunk,
plus the final text, on 8 LibriSpeech clips through the live path (and on 8 + 40 clips, 169 chunks, when the port
moved to the P4X); the search's tokens on 140 clips with phrase biasing on; and, in the timing runs of
Section~\ref{sec:speed}, the encoder CRCs of all eleven configurations in all of their runs. Everything matches. We therefore report
the PC build's WERs as the chip's. This is what lets us evaluate on 2620 and 2939 clips: the whole of test-clean runs
in about 15 minutes on a laptop in 16 processes, while the chip, at 0.7 times real time, would need about four hours
for test-clean alone.

\section{Speed}
\label{sec:speed}

\subsection{The history}

The encoder went from 51.3\,s per 1.28\,s chunk to about 0.6\,s. Table~\ref{tab:speed} lists each step that changed
the time, as it was measured on the chip during development. The first column is the first port, which kept float
glue between the int8 products and finished at 1.28\,s, still not real time. The second is the integer port of
Section~\ref{sec:integer}, which reached 630\,ms in one day. The stable version we kept dropped one of those steps, the
int8 front end, after our tests showed it lost 16\,dB of signal-to-noise ratio that the WER had not revealed
(Section~\ref{sec:failed}); that version ran at 726\,ms, and the move to the P4X brought it to 592\,ms. These numbers
come from successive builds, two chip revisions and a code layout that changed along the way, so a single step's gain
in this table is only indicative.

\begin{table}[tbp]
\centering
\caption{The history of the port: encoder time per 1.28\,s chunk on the chip after each step, as measured during
development (ESP32-P4 v1.3 at 360\,MHz, then the P4X at 400\,MHz; greedy search).}
\label{tab:speed}
\small
\begin{tabular}{lr@{\hspace{2.5em}}lr}
\toprule
\multicolumn{2}{c}{First port (float glue)} & \multicolumn{2}{c}{Integer port} \\
\cmidrule(r){1-2}\cmidrule(l){3-4}
Step & ms & Step & ms \\
\midrule
plain float C & 51,300 & integer products & 1,257 \\
2/4-bit products on PIE & 21,400 & PIE int8 rounding & 1,127 \\
int8 front end and joiner, fast math & 5,720 & int8 caches & 1,064 \\
two cores & 4,600 & softmax in int8, once & 983 \\
key-major attention & 2,750 & int16 FF hidden layer on PIE & 888 \\
L2 cache 256\,KB, 128-byte lines & 2,350 & int16 residuals, PIE norm and bypass & 859 \\
activation tables & 1,870 & int8 front end & 752 \\
attention $\times$ values in int8 & 1,680 & integer convolution modules & 686 \\
depthwise kernel & 1,554 & non-linear attention & 668 \\
L1 and L2 autoload & 1,354 & hot code in internal RAM & 630 \\
radix-8 Hadamard & 1,315 & \emph{stable, without int8 front end} & 726 \\
int8 queries and keys & 1,281 & P4X, rebuild only & 602 \\
 & & P4X, PSRAM at 250\,MHz & 592 \\
\bottomrule
\end{tabular}
\end{table}

\subsection{Each design step, measured in one build}

To give each step a reliable cost, we measured the main design decisions again on the P4X, in a single firmware build
of the released engine with run-time switches (Table~\ref{tab:speed2}). Each configuration boots on its own, reads the
weights from the card and processes the same two LibriSpeech clips (13 chunks) three times; the list was run three
times, so each row is the mean of 9 runs. The standard deviation over the runs is at most 0.5\,ms. Every configuration's
encoder output matches, CRC for CRC, the PC build of the same configuration, and the released design's matches the
reference values of the released engine, so each row computes exactly what it claims.

\begin{table}[tbp]
\centering
\caption{The design steps measured again in one build on the P4X (400\,MHz, two cores unless stated): ms per
1.28\,s chunk, mean of 9 runs (standard deviation $\le 0.5$\,ms). Chunk: features, encoder and beam search (4
hypotheses, no phrases). Rows b to h add one step each; i, j and k change one thing of h.}
\label{tab:speed2}
\small
\begin{tabular}{llrrr}
\toprule
 & Design & Encoder & Chunk & $\times$ real time \\
\midrule
a & int8 products on PIE, float in between (the first port) & 1043.8 & 1176.9 & 0.92 \\
b & integer products, int8 attention caches and softmax rows & 776.5 & 909.7 & 0.71 \\
c & \quad + int16 feed-forward hidden layer & 721.4 & 854.2 & 0.67 \\
d & \quad + int16 residual stream & 700.7 & 834.4 & 0.65 \\
e & \quad + integer convolution modules & 645.9 & 778.8 & 0.61 \\
f & \quad + integer non-linear attention & 631.2 & 763.8 & 0.60 \\
g & \quad + int8 self-attention values & 629.1 & 762.6 & 0.60 \\
h & \quad + integer attention weights: the released design & \textbf{620.3} & \textbf{753.5} & \textbf{0.59} \\
\midrule
i & h on one core & 979.5 & 1124.5 & 0.88 \\
j & h without the cache autoload & 665.9 & 821.9 & 0.64 \\
k & h with the screen refreshing (nothing drawn) & 630.6 & 769.2 & 0.60 \\
\bottomrule
\end{tabular}
\end{table}

From the first port's design to the released one, the encoder becomes 1.68 times faster (1043.8 to 620.3\,ms) and the
chunk 1.56 times (from 0.92 to 0.59 times real time). Integer products account for 63\,\% of that gain (267 of
424\,ms); the int16 feed-forward hidden layer and the integer convolution modules for 13\,\% each; the residual
stream, the non-linear attention and the attention weights for the rest. The second core makes the encoder 1.58
times faster, and the cache autoload saves 45.6\,ms of encoder time and 68.4\,ms of the chunk, since the search
benefits too. The order of the steps matters for how the gain is split: the int16 residual stream, for one, changes
the inputs of every module after it. Two configurations could not be run at all: plain C products and float products
need separate int8 or float copies of the weights and float caches, about 7.9\,MB that do not fit next to the 26.65\,MB
of weights.

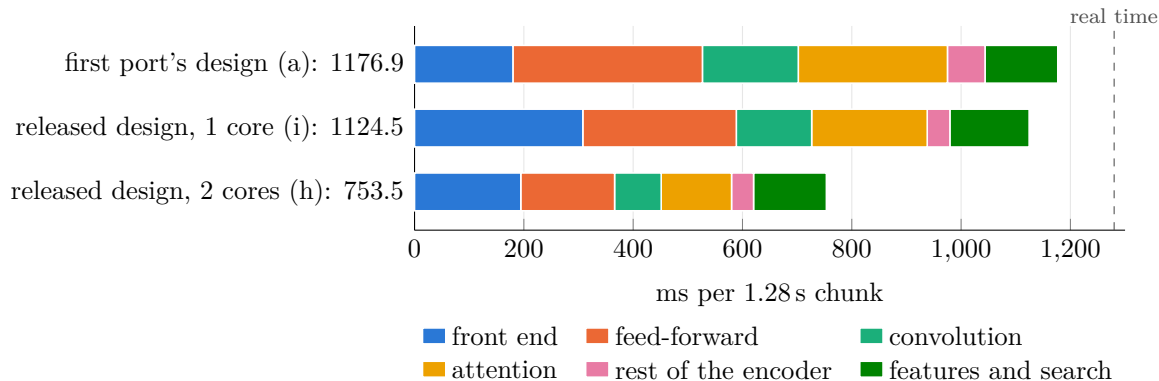
\begin{figure}[tbp]
\centering
\begin{tikzpicture}
\begin{axis}[
  xbar stacked, scale only axis, width=0.58\textwidth, height=2.7cm, bar width=5mm,
  xmin=0, xmax=1300, ymin=-0.6, ymax=2.6,
  ytick={0,1,2},
  yticklabels={{released design, 2 cores (h): 753.5}, {released design, 1 core (i): 1124.5},
               {first port's design (a): 1176.9}},
  xlabel={ms per 1.28\,s chunk},
  axis lines*=left, xmajorgrids, grid style={black!10},
  tick label style={font=\small}, label style={font=\small},
  every axis plot/.append style={draw=white, line width=0.6pt},
  legend style={at={(0.5,-0.42)}, anchor=north, legend columns=3, draw=none, font=\small,
                /tikz/every even column/.append style={column sep=8pt}},
  legend cell align=left, clip=false,
]
\addplot[fill=dvA] coordinates {(194.4,0) (307.9,1) (180.0,2)};
\addplot[fill=dvB] coordinates {(171.6,0) (280.8,1) (346.7,2)};
\addplot[fill=dvC] coordinates {(85.3,0) (137.9,1) (175.4,2)};
\addplot[fill=dvD] coordinates {(128.7,0) (211.4,1) (272.9,2)};
\addplot[fill=dvE] coordinates {(40.3,0) (41.5,1) (68.8,2)};
\addplot[fill=dvF] coordinates {(133.2,0) (145.0,1) (133.1,2)};
\legend{front end, feed-forward, convolution, attention, rest of the encoder, features and search}
\draw[dashed, black!70] (axis cs:1280,-0.6) -- (axis cs:1280,2.6);
\node[font=\scriptsize, anchor=south, black!70, inner sep=1pt] at (axis cs:1280,2.6) {real time};
\end{axis}
\end{tikzpicture}
\caption{Where a chunk's time goes on the P4X, in ms (rows a, i and h of Table~\ref{tab:speed2}; core~0's wall time
per part; the totals are in the row labels). Feed-forward and convolution: those modules of the 16 layers. Attention:
the attention weights, the non-linear attention and the self-attention. Search: the beam search with 4 hypotheses.
The 2/4-bit products are counted inside the parts that use them.}
\label{fig:breakdown}
\end{figure}

\paragraph{Where the time goes.} In the released design (row h, Figure~\ref{fig:breakdown}), the front end takes
194\,ms and the 16 layers 403\,ms: the feed-forward modules 172, the convolution modules 85, the attention weights 56, the non-linear attention
47, the self-attention 26, the norms 4, and the remaining 14 the middle bypass and the conversion of the residual
stream to int16 and back at each layer's ends. The 2/4-bit products, counted inside those modules, take 339\,ms, 55\,\% of
the encoder, and core~0 waits 25\,ms in all for core~1 at the end of the split products. The front end, still partly
in float, is the largest part that has not been worked on. The released engine's 620\,ms is 28\,ms more than the
592\,ms of the last development tree; we did not track down the difference.

\paragraph{In the application.} The live transcriber firmware (\texttt{firmwares/live\_transcriber} in the
repository) is slower than the engine alone: over the 8 test clips fed through
its debug link, a chunk takes 895.8\,ms (features 14.0, encoder 742.2, search with phrases and transcript rules
138.1\,ms), 0.70 times real time. The screen's refresh alone costs 10.4\,ms of encoder time (row k); the rest comes from
the other tasks of the application, mainly the user interface drawing on core~1, which uses 45\,\% of that core, and
we did not separate them further. A 3.1-minute live run of the microphone kept a mean of 904\,ms and a maximum of
935\,ms per chunk, lost no audio and used constant memory.

\subsection{Habits that helped}

\paragraph{Probe before the WER.} Each change was checked with a fast per-tensor signal-to-noise probe (4 minutes) before
the full WER run (15 minutes). The probe caught two losses that the WER missed: the static int8 front end (from 33.5 to
17\,dB) and static int8 scales in the convolution modules ($-4.6$\,dB, fixed with dynamic power-of-two scales). The WER
run's own noise is about $\pm$30 words on test-clean, so a single run cannot separate small effects; when we stacked
several unconfirmed steps, finding the bad one took two hours of bisection.

\paragraph{One build for timing.} During development, code layout alone moved the total by about $\pm$12\,ms between
builds with the same arithmetic, which is why Table~\ref{tab:speed2} switches the design at run time inside one build
instead of comparing builds.

\section{Streaming, search and phrases}
\label{sec:search}

\paragraph{Search.} The search is k2's modified beam search: at most one token per encoder frame, hypotheses with
the same tokens merged, and the best one length-normalized at the end. With one hypothesis it gives the greedy
search's tokens bit for bit. A beam of 4 brings test-clean from 4.52\,\% to 4.34\,\% (about 100 fewer word errors) for
138\,ms per chunk on the chip; a beam of 8 gives 4.30\,\% on the PC, too small a gain for the extra search time.

\paragraph{Commits and words.} A token is committed when every hypothesis agrees on it. A token still open after 75
frames (3\,s) is committed from the best hypothesis, and the hypotheses that disagree are dropped; without that rule a
long ambiguous stretch could hold words back indefinitely. A word is final when the next word starts or after 0.5\,s
without a token, and a pause of 1.5\,s starts a new line. We measure a word's delay in audio time, from the end of its
last token's frame to the moment it becomes final: the median is 1.9 to 2.3\,s depending on the material, and 95\,\% of
words are final within 4.1\,s.

\paragraph{Phrases.} A text file lists words and sentences to listen for; a word that the model spells differently
gets its spoken forms (\texttt{ESP32 = ESP thirty two}). Each phrase becomes token sequences in a trie: each word
split every way the model's pieces allow, the splits with the fewest pieces kept, and the first word also capitalized.
A hypothesis that follows a phrase gets a bonus of 2.0 per token, added to its log-probability; it loses the bonus if
the phrase breaks off and keeps it once the phrase is complete. The spoken forms are then rewritten to the written form in the transcript.

We tested this with the default list (4 words and 10 sentences) on 140 text-to-speech clips, 35 sentences in 4
voices (Table~\ref{tab:phrases}): 10 sentences contain words of the list, 15 contain its sentences, and 10 contain
neither but similar-sounding words. With the
phrases on, 46 of the 60 listed words are written as listed instead of 0, and the WER falls from 19.5\,\% to 3.4\,\%.
There are 3 false alarms (a listed phrase written that was not said), none of them in the 10 sentences without
phrases. On LibriSpeech test-clean, with the same list on, the WER moves from 4.34\,\% to 4.37\,\%.

\begin{table}[tbp]
\centering
\caption{Phrase biasing on 140 text-to-speech clips (beam 4). Words: the list's words written as listed. False alarms:
a listed phrase written but not said, over all clips (in the 40 clips without listed phrases).}
\label{tab:phrases}
\small
\begin{tabular}{lrrrr}
\toprule
 & Words & Sentences & False alarms & WER \\
\midrule
Phrases off & 0 / 60 & 62 / 72 & 0 (0) & 19.5\,\% \\
Phrases on (the chip) & 46 / 60 & 70 / 72 & 3 (0) & 3.4\,\% \\
\bottomrule
\end{tabular}
\end{table}

\section{Results}
\label{sec:results}

\subsection{LibriSpeech}

Table~\ref{tab:wer} gives the WERs on all of LibriSpeech test-clean (2620 clips) and test-other (2939 clips), scored
with Whisper's English text normalizer. The released model was run with sherpa-onnx 1.13.8 in streaming mode with the
same beam, with 2\,s of silence appended to each clip so that its last chunk is decoded; with less padding it loses the
last words and the comparison is unfair to it. Without the normalizer our test-clean numbers are 4.78\,\% greedy and
4.61\,\% beam.

\begin{table}[tbp]
\centering
\caption{WER on whole test sets, Whisper's English normalizer. Ours: the PC build, which computes what the chip
computes. Released: Kroko's int8 model in sherpa-onnx. Gap: ours minus released at beam 4, with a 95\,\% interval from
a paired bootstrap over clips where computed. The last row is a diagnostic: every FLEURS clip scaled to a peak of
0.5.}
\label{tab:wer}
\small
\setlength{\tabcolsep}{5pt}
\begin{tabular}{lrrrrrl}
\toprule
 & & \multicolumn{2}{c}{Ours (the chip)} & \multicolumn{2}{c}{Released} & \\
\cmidrule(lr){3-4}\cmidrule(lr){5-6}
Set & Clips & greedy & beam 4 & greedy & beam 4 & Gap, beam 4 \\
\midrule
LibriSpeech test-clean & 2620 & 4.52 & 4.34 & 4.03 & 3.94 & +0.40 \\
LibriSpeech test-other & 2939 & 10.89 & 10.26 & -- & 9.17 & +1.09 \\
FLEURS English (US) test & 647 & 17.75 & 15.95 & 14.92 & 13.68 & +2.26 [1.63, 2.95] \\
Common Voice 17.0 English test & 1000 & 20.72 & 19.56 & 18.85 & 18.47 & +1.09 [0.50, 1.69] \\
\midrule
FLEURS, scaled to peak 0.5 & 647 & 11.42 & 10.97 & 10.28 & 9.99 & +0.97 [0.68, 1.28] \\
\bottomrule
\end{tabular}
\end{table}

\subsection{Real speech}
\label{sec:realspeech}

LibriSpeech is read audiobooks, recorded and segmented with care. We also measured two public sets recorded in other
conditions (Table~\ref{tab:wer}): the whole English (US) test split of FLEURS~\cite{conneau2023fleurs}, 647 clips
and 1.77 hours, and 1000 clips drawn at random from the English test split of Common Voice
17.0~\cite{ardila2020commonvoice}, 981 speakers and 1.65 hours, resampled from 48\,kHz. Both are read or prompted
speech without added noise. The calibration audio of Section~\ref{sec:rounding} comes from FLEURS's training split,
not its test split. As before, our numbers come from the PC build and the released model's from sherpa-onnx, on the
same clips.

Both models make three to five times more errors here than on LibriSpeech, and the gap between them grows: 2.26 points
on FLEURS and 1.09 on Common Voice at beam 4, against 0.40 on test-clean, and the bootstrap intervals exclude zero on
both sets. The GPU simulator tells where the gap comes from. With the released weights it reproduces sherpa-onnx's
greedy WER (14.92\,\% on FLEURS, 18.90 against 18.85\,\% on Common Voice); with the 2/4-bit weights and float
activations it gives 17.78\,\% and 20.94\,\%, and the chip, with integer activations, 17.75\,\% and 20.72\,\%. The whole
gap comes from the weights. The integer arithmetic adds nothing here either.

Most of the FLEURS gap has a simple cause: the recordings are very quiet. 343 of the 647 clips peak below 0.01 of full
scale ($-40$\,dBFS), and the median clip's RMS level is $-60$\,dBFS. On such clips both models tend to miss the start of
the sentence, ours more often: the words lost before the first recognized word are 12.8\,\% of the reference words
for ours (greedy) and 9.2\,\% for the released model. Scaled to a peak of 0.5, these leading deletions almost vanish
(from 1042 to 51 words for ours, from 725 to 17 for the released model), both WERs fall by 3.7 to 6.3 points, and the
gap shrinks to 0.97 points at beam 4, about 10\,\% relative, the same proportion as on LibriSpeech. On the board the
input level is set by the microphone gain, so this sensitivity matters less there than on recordings, but it is real.

On Common Voice, at normal levels, our extra errors also sit mostly at the start of the clips: our model drops more
leading words than the released one in 93 clips and fewer in 10. Adding one second of silence before every clip
changes neither WER, so the cause is not the cold start of the stream; we have not found it. A beam of 4 recovers some
of these first words, which is why it helps our model more than the released one (1.16 against 0.37 points on Common
Voice). Names are hard for both models (42 to 47\,\% of the name-like words in Common Voice are wrong); that is Kroko's
own weakness, not the port's.

\subsection{Memory}

Table~\ref{tab:memory} shows where the memory goes. The live transcriber's release build leaves 121\,KB of internal
RAM free, and its debug build, with a USB link to a PC, 73\,KB; the repository's main firmware, which adds
voice-control demos to the transcriber, leaves 104\,KB. Internal RAM was tight during the port: with the 256\,KB L2 cache
about 70\,KB were left for everything else, and a build with 40\,KB free failed at boot when ESP-IDF reserved its
internal DMA pool.

\begin{table}[tbp]
\centering
\caption{PSRAM use while listening (32\,MB).}
\label{tab:memory}
\small
\begin{tabular}{lr}
\toprule
Part & Size \\
\midrule
weights, used in place & 26.65\,MB \\
engine state and workspace & about 4\,MB \\
screen frame buffer ($1024\times600$, RGB565) & 1.2\,MB \\
free, at the lowest & about 0.5\,MB \\
\bottomrule
\end{tabular}
\end{table}

\subsection{Comparison}

Table~\ref{tab:compare} puts the numbers next to other small recognizers. The comparison has to be read with care.
Oído's WERs, like ours, come from a PC build of its engine, with its own text normalization; its best streaming
numbers use a 1.3M-parameter language model. Moonshine's and Whisper's numbers are from the Moonshine paper, measured
on a PC. The chips differ: one 240\,MHz Xtensa core for Oído, two 400\,MHz RISC-V cores for us, and Oído's model is
five times smaller and fits a chip with 8\,MB of PSRAM. With those caveats, our streaming WERs are below the offline
ones of Moonshine Tiny and Whisper tiny.en and above those of Oído's whole-utterance CTC model. Against Oído's
streaming variant they are 1.8 and 2.2 points lower greedy against greedy, and 0.5 and 0.8 points lower with our beam
search against its language model. Ours is the only open-vocabulary entry in the table that is both streaming and
faster than real time on a microcontroller.

\begin{table}[tbp]
\centering
\caption{Other small speech recognizers on LibriSpeech, as published by their authors (ours: measured here). WER in
\%. RT: processing time over audio duration; ours is the live firmware with its screen (0.59 for the engine alone,
Table~\ref{tab:speed2}); Oído's streaming RT is for 1.28\,s chunks with its language model running. ESP-SR's MultiNet~\cite{espsr}, which recognizes a
fixed list of up to 200 commands on the ESP32-S3, does not transcribe open speech and is left out.}
\label{tab:compare}
\footnotesize
\setlength{\tabcolsep}{4pt}
\begin{tabular}{lrlllrrr}
\toprule
System & Params & Weights & Runs on & Streaming & clean & other & RT \\
\midrule
This work, beam 4 & 65.7M & 2/4-bit, 26.65\,MB & ESP32-P4, 2 cores & yes & 4.34 & 10.26 & 0.70 \\
This work, greedy & 65.7M & 2/4-bit, 26.65\,MB & ESP32-P4, 2 cores & yes & 4.52 & 10.89 & \\
Kroko as released, beam 4 & 65.7M & int8, 71\,MB & PC & yes & 3.94 & 9.17 & \\
Oído CTC, greedy~\cite{varela2026oido} & 13M & int8, 14.0\,MB & ESP32-S3, 1 core & no & 3.70 & 8.23 & 1.97 \\
Oído streaming, greedy~\cite{varela2026oido} & 13M & int8 & ESP32-S3, 1 core & yes & 6.27 & 13.06 & \\
Oído streaming, with LM~\cite{varela2026oido} & 14.3M & int8 & ESP32-S3, 1 core & yes & 4.86 & 11.03 & 1.96 \\
Moonshine Tiny~\cite{jeffries2024moonshine} & 27.1M & float & PC & no & 4.52 & 11.71 & \\
Whisper tiny.en~\cite{jeffries2024moonshine,radford2023whisper} & 37.8M & float & PC & no & 5.66 & 15.45 & \\
\bottomrule
\end{tabular}
\end{table}

\section{Toward 16 MB}
\label{sec:push}

Once the model ran, we tried to make it much smaller. The goal was a memory budget, not speed: Kroko together with a
second model in the same 32\,MB of PSRAM, which needed Kroko at about 16\,MB with every matrix at 2 bits. Post-training
rounding alone could not get there (every large matrix at 2 bits with the full method of
Section~\ref{sec:rounding}: 13.40\,\%, Table~\ref{tab:ablation}), so this attempt used training, in two steps.

\paragraph{Slimming.} We cut stack 3's feed-forward modules, which hold 18.9M of its 30.5M parameters. Each hidden unit
was scored by the root-mean-square of its SwooshL activation times the norm of its output weights, on 800 clips; the
global top half was kept, with widths rounded to multiples of 128 (at least 384). This took the model from 65.75M to
56.31M parameters. The cut units were faded out over the first 30\,\% of the training steps, and stacks 3 to 5 were
trained to reproduce the original encoder output (a normalized squared error) on 12.4 hours of LibriSpeech dev-clean
and FLEURS English. The slim model scores 4.29\,\% on test-clean against 4.02\,\% for the original, both in float in
the PyTorch pipeline used for this study. Dropping a whole layer of stack 3 instead, without training, gave 4.13\,\%
for 58.13M parameters; we chose the narrower feed-forwards for the larger saving.

\paragraph{Quantization-aware training.} We then made the GPU simulator of Section~\ref{sec:integer} trainable: it
streams chunk by chunk, rounds the weights to the chip's format with a straight-through
gradient~\cite{bengio2013ste}, and reproduces the chip's integer activations. It matched the PyTorch model at 92 to
94\,dB once a bug was fixed (it had rounded the attention's position projections to 2 bits). With every matrix at
2 bits, the WER on the study's monitoring clips was 45.5\,\% before training and 9.6 to 13.1\,\% after, so we kept 113 matrices (7.85M parameters) at 4
bits, chosen by sensitivity for an 18\,MB budget. The weights were blended from float to rounded over 30\,\% of 1480
steps, then the integer activations and an int8 front end were added for 100 steps each.

\paragraph{Result.} The final model, measured, is 19.63\,MB and scores 5.63\,\% on test-clean, with float or integer
activations alike: 1.35 points above the slim float model, for 7\,MB less than our published file. It was never run on
the chip. The 16\,MB target was not met: the estimate of 16.03\,MB assumed every matrix at 2 bits and nothing else,
and the 4-bit matrices, the float16 small tensors, the int8 decoder and joiner and the smaller groups brought the
real file to 19.63\,MB. A storage format with 8-bit log norms, int8
small tensors and a 2/4-bit decoder would bring the same model to about 17.4 to 17.6\,MB, but that is a projection that
needs a new training run. We stopped there and kept the capacity for an Arabic model, where it is needed more.

Two observations from this attempt. Short quantization-aware training on 12 hours of audio did not beat post-training
rounding where both apply: starting the training from the GPTQ-rounded model instead of the plain one helped before
training (7.23\,\% against 10.57\,\% on a small monitoring set) and not at all after it. And the int8 front end, which lost
16\,dB untrained on the chip, was lossless after training in the simulator.

\section{What did not work}
\label{sec:failed}

\begin{itemize}
\item \emph{Plain 2 bits.} Rounding every large matrix to 2 bits to the nearest level gives 21.76\,\% on test-clean
against 4.02\,\% for the unquantized model (Table~\ref{tab:ablation}). Errors compound: during development, the
single-matrix errors summed to a predicted 0.126, and the measured error was 0.196.
\item \emph{Short quantization-aware training at 2 bits.} Thirty minutes of self-distillation (7.4 hours of FLEURS)
reached a lower error on FLEURS than on LibriSpeech (0.02 against 0.053) and lost to GPTQ without training (27.2\,\%
against 25.4\,\% on the author's recordings).
\item \emph{The correction applied to a whole layer at once.} It counts upstream error several times: error 0.33
instead of 0.034.
\item \emph{4 bits for the decoder and joiner.} From 10.1\,\% to 16.6\,\% on the author's recordings; they stay int8.
\item \emph{3 bits.} The best 2/3/4-bit allocation had about 8\,\% less error than the 2/4-bit one at the same size,
while raising the 2/4-bit budget from 22 to 25\,MB halved the error. Not worth a third decode kernel.
\item \emph{Vector codebooks.} A 4-dimensional codebook of 256 entries reduced the error per weight by 17\,\% and the
model's by 11\,\%, worth about 0.5 to 1\,MB; we stopped there.
\item \emph{Replacing the largest feed-forward.} Distilling stack 3's largest feed-forward (1.97M parameters) into a
Hadamard-structured MLP or into a pruned copy never came close: 2.8 to 3.7\,dB for the structured MLPs against 4.9\,dB
for a plain linear map, and 12\,dB at half the hidden width after fine-tuning.
\item \emph{All-int8 activations} (12.75\,\% against 7.27\,\%), \emph{static int8 inputs to the products} (26.1 against
37.8\,dB) and \emph{a static int8 front end} ($-16$\,dB, invisible in the WER run).
\item \emph{On the chip:} an L2 cache of 512\,KB (the chip does not boot), 256\,KB with 64-byte lines (it hangs at
initialization), manual L2 preloading (1120 to 1183\,ms), and loop unrolling by the compiler.
\end{itemize}

\section{Limitations}
\label{sec:limits}

We did not measure energy or power; we had no instrument for it. The system is English only. The test sets are read
or prompted speech: LibriSpeech, FLEURS, Common Voice, and text-to-speech clips for the phrases. Noisy, far-field and
conversational speech were not measured, and on two noisy recordings of the author both the released model and ours
made 43\,\% errors, so noise robustness is an open question that belongs to the model rather than to the port. The
WERs come from the PC build; we checked bit-exactness on 48 LibriSpeech clips, all kernels and 140 phrase clips, not
on whole test sets on the chip. The 2/4-bit weights cost half a point on test-clean, about one point on test-other and
Common Voice, and two on FLEURS as published; the model with these weights is more sensitive to very quiet input and
drops first words more often than the released one, for reasons we have not established. The firmware needs chip revision v3 (the P4X) and 32\,MB of PSRAM, and the weights need a
microSD card. Finally, the method was developed on one model; we have not shown that the same settings transfer to
other models.

\section{Conclusion}

A 65.7M-parameter streaming transducer runs on the ESP32-P4 at 0.70 times real time with a beam of 4, at 4.34\,\% WER
on LibriSpeech test-clean. Getting there needed no training: a sensitivity-based choice of 2 or 4 bits per matrix,
a sequential rounding that corrects each matrix toward the original model, integer activations chosen by measurement,
and a chip that is bit-exact with its PC build so that every number could be checked on whole test sets. The integer
arithmetic turned out to cost nothing on every set we measured; the cost is in the weights, half a point on
LibriSpeech and one to two points on harder speech. Closing that gap, with a larger bit budget or with training on
much more audio than the 12 hours of Section~\ref{sec:push}, and moving the rest of the front end to integers are the
next improvements. Beyond them, we plan to support the ESP32-S31, which has the same PIE 2.2 instructions, and to
build a model for Modern Standard Arabic.

\paragraph{Availability.} The engine, the firmware, the evaluation scripts and the LibriSpeech transcripts behind the
WERs are at \url{https://github.com/BoumedineBillal/esp32-pie-asr} (Apache-2.0). The weights are at
\url{https://huggingface.co/BoumedineBillal/esp32-pie-asr-kroko-en} under CC-BY-SA 4.0, as Kroko's.

\paragraph{Acknowledgments.} Thanks to Banafo for Kroko, to Cactus Compute for documenting their weight format, to the
k2-fsa developers for icefall and sherpa-onnx, and to Espressif for ESP-IDF.

\bibliographystyle{plainnat}
\bibliography{references}

\end{document}